\documentclass[12pt]{article}
\usepackage{epsf,epsfig,psfig,float,amssymb,latexsym}
\usepackage{graphics,psfrag}
\newcommand{\be}{\begin{equation}}
\newcommand{\ee}{\end{equation}}
\newcommand{\ba}{\begin{eqnarray}}
\newcommand{\ea}{\end{eqnarray}}

\newcommand{\nn}{\nonumber}

\newcommand{\vs}{\vspace{-0.20cm}}

\begin{document}

\thispagestyle{empty}

\vspace{2cm}

\begin{center}
{\Large{\bf Final State Interactions in Hadronic $D$ decays}}
\end{center}
\vspace{.5cm}

\begin{center}
{\Large Jos\'e A. Oller\footnote{email: oller@um.es}}
\end{center}

\begin{center}
{\it {\it Departamento de F\'{\i}sica. Universidad de Murcia.\\ E-30071 
Murcia, Spain.}}
\end{center}
\vspace{1cm}

\begin{abstract}
\noindent
We show that the large  corrections due 
to final state interactions (FSI) in the  $D^+\rightarrow \pi^-\pi^+\pi^+~,$
$D^+_s\rightarrow \pi^-\pi^+\pi^+~,$ and $D^+\rightarrow K^-\pi^+\pi^+$ decays can be
accounted for by invoking scattering amplitudes in agreement with those derived 
from phase shifts studies. In this way, broad/overlapping resonances in S-waves
are properly treated and the phase motions of the transition amplitudes are
 driven by the scattering matrix elements determined in many other
 experiments. This is an important step forward in resolving the puzzle 
 of the FSI in these decays. We also discuss why the $\sigma$ and $\kappa$ resonances, hardly 
 visible in scattering experiments, are much more prominent 
 and clearly visible in these decays without destroying the agreement with 
 the experimental $\pi\pi$ and $K\pi$ low energy S-wave phase shifts. 
\end{abstract}

\vspace{2cm}


\newpage

\section{Introduction}
\label{sec:intro}
\def\theequation{\arabic{section}.\arabic{equation}}
\setcounter{equation}{0}
During the last few years a series of theoretical works that combine the Chiral Perturbation Theory (CHPT)
 expansion \cite{wein,gl} with unitarity in a consistent way \cite{npa,nd,iamdoba,iamprl,hannah,jamin,mixing}, order by 
 order,  have showed up 
the close relationship between the leading order CHPT amplitudes (driven by the chiral structure and the actual value 
of the order parameter for the spontaneous breakdown of chiral symmetry, $f_\pi$) and the presence, mass and 
width, together with other properties, of the lightest scalar resonances. These results, concerning the $\sigma$ resonance,
 have been also confirmed by the solution of the Roy equations \cite{roygas}. 

Also recently, but from an experimental point of view, charmed decays have revealed as a powerful source for the
study of the properties of the scalar resonances offering experiments with high statistics, which is of foremost 
importance taking into account the traditionally lack of   high statistics experiments
in the scalar sector. In this respect, the studies of the Fermilab E791 Collaboration
 concerning the  decays $D^+\rightarrow \pi^-\pi^+\pi^+$ \cite{prld3pi}, 1686 events, and 
 $D^+\rightarrow K^-\pi^+\pi^+$ \cite{prldk2pi}, 15090 events, were the first to point out 
 statistically significant contributions of the $\sigma$ and $\kappa$ resonances, respectively.
 Other Collaborations on $D$ decays also report with high statistical significance  the existence of a 
 $\sigma$ resonance from the $D^0\rightarrow  K^0_s\pi^+\pi^-$ decays, like the CLEO Col. \cite{cleo}, 
 $5299\pm 73$ events,  Belle Col. \cite{belle}, 57800 events, and the BaBar Col. \cite{babar},
 81396 events. 
  Thus, we can conclude that
nowadays, with the new data from charmed decays, the existence of the $\sigma$ should be taken for grant. 

Despite these positive facts concerning the charmed decays, from the theoretical point of view several questions
arise regarding the analyses followed in the previous references, particularly in the S-waves. We concentrate
 in this work in the results of the
E791 Collaboration with respect to the $D^+_s\rightarrow \pi^-\pi^+\pi^+$ \cite{prlds3pi}, 
$D^+\rightarrow \pi^-\pi^+\pi^+$ \cite{prld3pi}
and $D^+\rightarrow K^-\pi^+\pi^+$ \cite{prldk2pi} decays. Although the final state is a three body one, we would 
expect that, at least, at the low energy tail of the invariant mass of the neutral $\pi^+\pi^-$ subsystem,
 where the $\sigma$ is observed, 
the interactions of this two pion subsystem with the other pion should be quite soft due to the largeness
of the $D^+$ mass. In this way, the movement of the phase of the $\sigma$ resonance across the Dalitz plot 
should be given, according to Watson final state theorem, by the isospin ($I$) 1/2 S-wave $\pi\pi$ phase shifts.
However, the Breit-Wigner (BW) employed for the $\sigma$ meson in ref.\cite{prld3pi}, as it is well known, 
does not fulfill this property, as we discuss below in more detail. Analogous comments are also appropriate 
 for the $\kappa$ resonance in the $K^-\pi^+\pi^+$ decay \cite{prldk2pi} for the $K^-\pi^+$ subsystem 
and the $K\pi$ $I=1/2$ S-wave phase shifts. In addition, controversial properties were also reported by
 this collaboration regarding the $f_0(980)$ and 
the $K^*_0(1430)$ resonances. We will develop in this work alternative parameterizations, based on CHPT plus 
unitarity (chiral unitary approach \cite{review,kn,nn}),  to those employed by the E791 Collaboration that 
do not show
 the mentioned  problems with the S-waves but also contain the $\sigma$ and $\kappa$ resonance poles. 
  We will also discuss, based on the presence of Adler zeros, why large destructive non-resonant
   contributions (called backgrounds by theorists) that completely distort the peak shapes of the 
   $\sigma$ and $\kappa$ resonances are present
 in $\pi\pi$ and $K\pi$ S-wave  scattering at low energies, respectively.  We will also offer a reason why 
 these backgrounds are not present in charmed decays (also applicable to $B$ decays) due to the lack 
 of such zeroes occurring only at a specific energy. As a result, we will put on 
 sounder theoretical grounds the important findings of the E791 Collaboration about the relevant
  role played by the $\sigma$ and $\kappa$ mesons.\footnote{Also called $f_0(600)$ and $K^*_0(800)$ in the PDG \cite{pdg04}, respectively.}
    Let us stress that in this work we focus on the solution
  of the previous problems associated with the S-waves but we do not attempt to account for full
   three body unitarity, this is beyond the scope of this work.
 
The content of this paper is as follows. After this brief introduction, we consider the 
$D^+\rightarrow \pi^-\pi^+\pi^+$, $D^+_s\rightarrow \pi^-\pi^+\pi^+$ and
 $D^+\rightarrow K^-\pi^+\pi^+$ decays in sections \ref{sec:3pi}, \ref{sec:ds3pi} and 
 \ref{sec:dk2pi}, respectively. We review the formalism and main findings of 
 refs.\cite{prld3pi,prlds3pi,prldk2pi} regarding these decays, in order, and 
develop our formalism  to take into account  FSI in section \ref{sec:3pi}   
and apply it to the previous three decays. We also discuss why one should expect clear, although broad, 
peak structures associated with the $\sigma$ and $\kappa$ resonances in these decays. We end with section 
\ref{sec:con} pointing out the most relevant results of our study.

\section{FSI in the $D^+\rightarrow \pi^- \pi^+\pi^+$ Decay}
\label{sec:3pi}
\def\theequation{\arabic{section}.\arabic{equation}}
\setcounter{equation}{0}

\begin{figure}[ht]
\psfrag{degrees}{\small Phase shifts (degrees)}
\psfrag{G}{\small GeV}
\psfrag{absolute}{\small (Absolute value)$^2$}
\psfrag{Kpi}{\small $K\pi$ $I=1/2$ S-wave}
\centerline{\epsfig{file=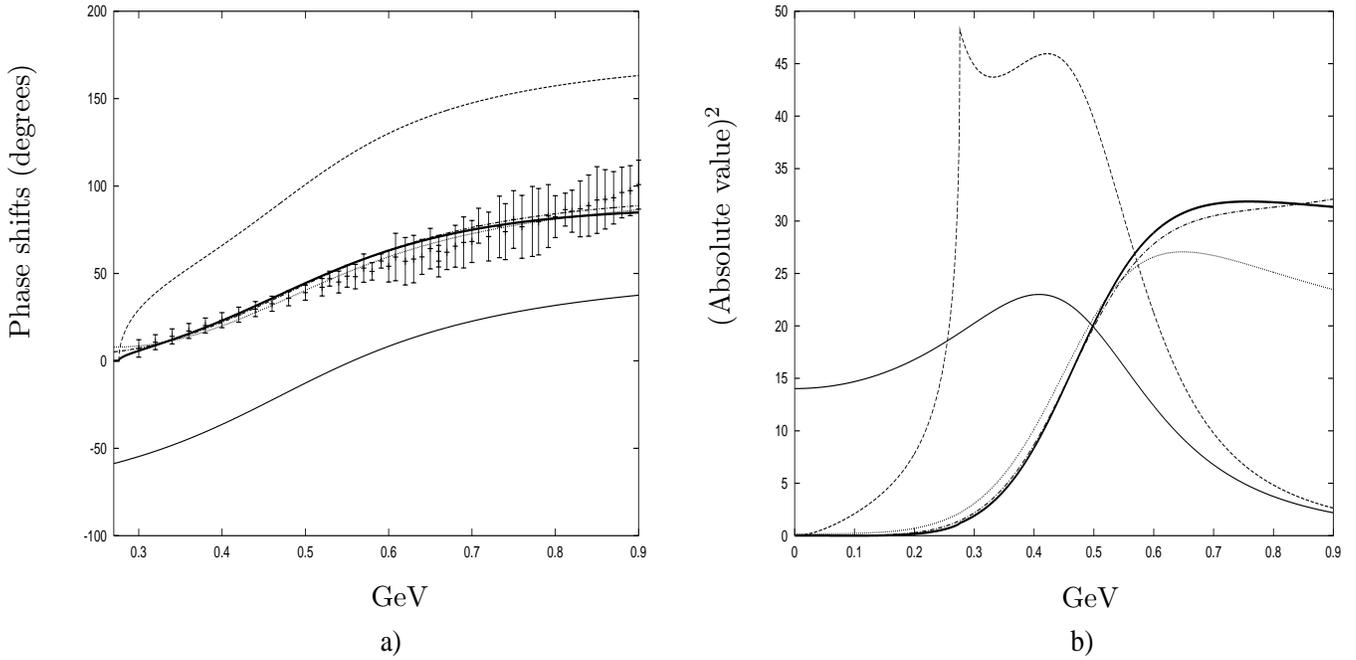,height=3.5in,width=7.0in,angle=0}}
\vspace{0.2cm}
\caption[pilf]{\protect \small S-wave $I=0$ $\pi\pi$ phase shifts (left panel) 
and modulus square of this partial wave  (right panel), normalized such that the residue of 
the $\sigma$ pole is one. The data points are the elastic S-wave $I=0$ $\pi\pi$ phase 
shifts from several experimental references and  the errors show the variation in the 
 values from one set to another, as explained in detail in ref.\cite{nd}. 
The dashed lines correspond to the BW of the $\sigma$ resonance 
employed by the E791 Collaboration \cite{prld3pi}. 
The thick solid lines are the full results of ref.\cite{npa} when keeping only the $\pi\pi$ channel 
while the thin ones correspond to the pure $\sigma$ pole contribution of Eq.(\ref{3pitaylor}). 
 The phase shifts and absolute value from the contribution of the $\sigma$ pole plus the first 
 non-resonant term in Eq.(\ref{kpitaylor}) are shown by the dotted lines. Finally, the 
dashed-dotted lines are the results from the Laurent expansion of 
Eq.(\ref{3pitaylor}) keeping all the terms shown.
\label{fig:3pitaylor}}
\end{figure}

The study of the  Dalitz plot of the decay 
$D^+\rightarrow \pi^- \pi^+\pi^+$, with a sample of 1686 candidates,
 was performed in ref.\cite{prld3pi} within the Fermilab E791 Collaboration. The estimated
signal to background ratio is 2:1, hence we are
 left with 1124 events and in the study of this decay we will normalize
  our results to this number of events. Previous studies with lower
 statistics can be also found in this reference from the E687 Collaboration. 
 In order to introduce the controversial situation regarding the amplitude employed 
 for describing data in ref.\cite{prld3pi}, let us briefly discussed the fitting process
 followed by ref.\cite{prld3pi}. In this reference, when  the decay amplitude is modeled 
 as the coherent sum of well established resonances \cite{pdg04}, following the 
 isobar model, the non-resonant term is dominant
 while the description of the fit is poor. Indeed the $\chi^2$ per degree of freedom, 
 $\chi^2/\nu$, turns out to be 254 for 162 degrees of freedom \cite{prld3pi}. The main 
 discrepancies between data and the parametrized amplitude come, particularly, from the 
 low energy region, below 0.5 GeV$^2$. In order to improve the quality of the
fit of the Dalitz plot the authors of ref.\cite{prld3pi}
 included another $\pi^+\pi^-$ resonance corresponding to the $\sigma$ 
 resonance whose 
 mass and width were allowed to float in the fit. The quality of the fit substantially 
improves with a $\chi^2/\nu=138/162$ and finds values of $478\pm 24$  and  $324\pm 42$ MeV for 
the mass and width of the $\sigma$, respectively. However,  
the phase of the Breit-Wigner (BW) employed to describe the $\sigma$ resonance does not
 follow the elastic S-wave $I=0$ $\pi\pi$ phase shifts. This is shown in 
 Fig.\ref{fig:3pitaylor}a by the dashed line in comparison with the data points corresponding 
 to the elastic S-wave $I=0$ $\pi\pi$ phase shifts
 from several experimental references. Indeed, the errors show the variation in the 
 values from one reference to another, see ref.\cite{nd} for more details. The discrepancy 
 is manifest, specially close to threshold where the variation 
 of the phase for the BW is much faster than that corresponding to data, despite that at the shown 
 energy the $\pi\pi$ interaction is elastic since the $K\bar{K}$ is above 1 GeV and 
 the $4\pi$ only starts contributing in a significant way above around 1.3-1.4 GeV. 

Denoting by 1 the $\pi^-$ and by 2 and 3 the equal positively charged pions, 
the amplitude employed in ref.\cite{prld3pi} can be written as:
\be
{\cal A}=a_0 e^{i\delta_0} {\cal N}_0+\sum_{n=1}^N a_n e^{i\delta_n} {\cal
A}_n(s_{12},s_{13}) {\cal N}_n~.
\label{expprld3pi}
\ee
We describe in detail the ingredients present in the previous equation since 
we found several errata and omissions in ref.\cite{prld3pi} that are necessary in order 
 to reproduce their amplitudes. In Eq.(\ref{expprld3pi})
 the first term is the non-resonant contribution and the 
other ones originate from the exchange of resonances. Every resonant contribution is 
Bose symmetrized for the equally charged pions, ${\cal A}_n={\cal A}_n[(12)3]+{\cal A}_n[(13)2]$, 
 as usual. The parenthesis around 12 mean that the particles 1 and 2 form the resonant state, and 
analogously for (13). The coefficients $a_n$ and $\delta_n$, for $n\geq 0$, are real
 constants that float in the fit, the $a_n$ 
are magnitudes and the $\delta_n$ phases. Finally, the ${\cal N}_n$, 
$n\geq 0$, are normalization factors\footnote{Omitted in the description 
of the amplitudes given in ref.\cite{prld3pi}, but used in their results.} 
given by:
\be
{\cal N}_n=1/\left(\int ds_{12} ds_{13} |{\cal A}_n(s_{12},s_{13})|^2\right)^{1/2}~.
\label{normalization}
\ee
The ${\cal A}_n[(12)3](s_{12},s_{13})$ amplitudes for $n>1$ correspond to 
the product 
\be {\cal
A}_n[(12)3](s_{12},s_{13})=BW_n(s_{12}) F_D^{(J)}(s_{12})
F_n^{(J)}(s_{12}) {\cal M}_n^{(J)}[(12)3]~.
\ee
 The Breit-Wigner propagator is given by,
\ba
BW_n(s_{12})&=&\left(s_{12}-m_n^2+i m_n \Gamma_n(s_{12})\right)^{-1}~,\nn\\
\Gamma_n(s_{12})&=& \Gamma_n(m_n^2)\frac{m_n}{\sqrt{s_{12}}}
\left(\frac{p_1}{\widetilde{p}_1}\right)^{2 J+1}
\frac{F_n^{(J)}(p_1)^2}{F_n^{(J)}(\widetilde{p}_1)^2}~,
\label{bwpropa}
\ea
with $m_n$ and $\Gamma_n(s_{12})$ the mass and 'energy dependent' width of the (12) 
 resonance and $p_1$ is the three-momentum of the particle 1 in the (12) rest frame. 
 Furthermore, $\widetilde{p}_1$ is $p_1$ evaluated at the resonance mass. 
 On the other hand, the ${\cal M}_n^{(J)}$ are angular dependent factors and read:
\ba
{\cal M}_n^{(0)}[(12)3]&=&1 ~,\nn\\
{\cal M}_n^{(1)}[(12)3]&=&-2 p_1 p_3 \cos\theta_{13}~,\nn\\
{\cal M}_n^{(2)}[(12)3]&=&\frac{4}{3} (p_1 p_3)^2(3 \cos^2 \theta_{13}-1)~.
\label{angularmn}
\ea
where $p_1$ and $p_3$ are the three-momenta of particles 1 and 3 and $\theta_{13}$ their 
relative angle, all of them referred to the (12) rest frame. The $F_D^{(J)}$ and $F_n^{(J)}$ 
are Blatt-Weisskopf penetration factors that depend on the spin $J$ of the resonant 
state, and are given by:
\ba
F_D^{(0)}&=&1~,\nn\\
F_D^{(1)}&=&1/\sqrt{\left(1+(r q_3)^2\right)}~,\nn\\
F_D^{(2)}&=&1/\sqrt{\left(9+3 (q p_3)^2+(r p_1)^4 \right)} ~,\nn\\
F_n^{(0)}&=&1~,\nn\\
F_n^{(1)}&=&1/\sqrt{\left(1+(r p_1)^2\right)}~,\nn\\
F_n^{(2)}&=&1/\sqrt{\left(9+3 (r p_1)^2+(r p_1)^4 \right)} ~,
\ea
with $q_3$ the three-momentum of particle 3 in the $D^+$ rest frame. 

The set of resonances that exchange in Eq.(\ref{expprld3pi}) contain the
$\rho^0(770)$, $f_0(980)$, $f_2(1270)$, $f_0(1370)$, $\rho^0(1450)$ and, in the 
parameterization that reproduces faithfully the experimental data, the $\sigma$. 
As discussed above the most controversial aspects present in 
 Eq.(\ref{expprld3pi}) involves the $I=0$ S-wave $\pi\pi$ partial wave.

Now, we want to show that we are able to reproduce the amplitude 
given in Eq.(\ref{expprld3pi}) as employed in ref.\cite{prld3pi} but 
using the S-wave $I=0$ $\pi\pi$, $K\bar{K}$ coupled channel partial 
waves derived in ref.\cite{npa}. These T-matrices were obtained from 
Chiral Perturbation Theory (CHPT) at leading order \cite{wein,gl} together with
 a unitarization scheme compatible with the chiral 
expansion (chiral unitariy approach). Furthermore, these matrices are not only able 
to reproduce the scattering data of the $I=0,$ 1 S-wave amplitudes up to around 1.2 GeV but 
also have been successfully tested by now in a vast number of production processes that pick up 
large corrections by FSI from these partial waves, see e.g. \cite{gama,fi,jpsi,bdecays}.

The S-wave $I=0$ T-matrices from ref.\cite{npa} contains two poles in the
appropriate Riemann sheets corresponding to the $\sigma$ and $f_0(980)$ resonances. 
Their pole positions are around $448-i\,224$ and $990-i 13$ MeV, respectively. 
Let us remark that the given poles are obtained in the full 
T-matrices derived in ref.\cite{npa}, which are not any sum of  pole contributions.  
These resonances originate because of the self interactions between the pseudoscalars and 
are of dynamical origin, disappearing in the large $N_c$ limit \cite{nd,ramonet}. This is why 
they are said to be of dynamical origin.

Due to the already mentioned discrepancy between phase shifts and the phase motion in energy of the 
BW for the $\sigma$, let us perform a Laurent series around 
the $\sigma$ pole position of the $I=0$ $\pi\pi$ S-wave amplitude,
 \be
 t_{11}=\frac{\gamma_0^2}{s-s_\sigma}+\gamma_1+\gamma_2(s-s_\sigma)+\ldots~.
\label{3pitaylor} 
  \ee
 with the pole position and residua obtained from ref.\cite{npa} in the elastic case, with the 
 values:
 \ba
 s_\sigma&=&(0.47-i 0.22)^2 \hbox{~GeV}^2~,~\gamma_0^2=5.3+i 7.7 \hbox{~GeV}^2~,\nn\\
 \gamma_1&=&-8.1+i 36.9 ~,~\gamma_2=1.1+i0.1 \hbox{~GeV}^{-2}~.
 \label{residua_s}
 \ea
 In  Fig.\ref{fig:3pitaylor} we show in the left and right panels the phase and 
normalized absolute value of this partial wave, respectively. 
The thinner solid lines correspond to the pole contribution in Eq.(\ref{3pitaylor}). 
We see in Fig.\ref{fig:3pitaylor}a that the phase of this pole contribution does not vanish
 at threshold, but that 
runs parallel to the experimental phase shifts, so that the difference with respect to them 
 keeps constant along energy up to the increase in the last points due to the closeness 
 of the $f_0(980)$ and the opening of the $K\bar{K}$ channel. Thus, the phase of the 
 $\sigma$ pole contribution from Eq.(\ref{3pitaylor}) does follow the motion of
 the experimental S-wave $I=0$ $\pi\pi$ phase shifts, in contrast with the BW phase indicated 
 by the dashed line employed in ref.\cite{prld3pi}. It is also 
 worth stressing that the phase of the pure $\sigma$ pole contribution starts at $-58$ 
 degrees at threshold and this is the reason why its value is not $+90$ degrees at the mass 
 for the $\sigma$, $466$ MeV, but happens much later, around 1 GeV. By the same reason, 
 for $s\rightarrow \infty$ one gets only $+122$ degrees from this pole contribution instead 
 of the usual $+180$ degrees.  The agreement between the experimental 
phase shifts and the expansion (\ref{3pitaylor}) is reached rather fast. The dotted lines 
in Fig.\ref{fig:3pitaylor} correspond 
to keep as well the $\gamma_1$ term in Eq.(\ref{3pitaylor}), while the dashed-dotted ones correspond to keep all the terms
shown in this equation. The thick solid lines are the full results from ref.\cite{npa} in the 
elastic case, removing the $K\bar{K}$ channel. In Fig.\ref{fig:3pitaylor}b we consider the absolute value of the partial wave under consideration, 
but normalized such that the residue at the pole position is one, so that the series in Eq.(\ref{3pitaylor}) 
 is divided by $\gamma_0^2$. The source of each line is the same 
as already explained and let us notice that this figure starts at 0 GeV, below 
threshold.  We want to stress three important facts: i) 
The markedly different behaviour of the $\sigma$ BW employed by the E791 Collaboration and 
the absolute value from the partial wave of ref.\cite{npa}. ii) The first peak in the BW. This 
peaks results because the BW formula (\ref{bwpropa}) for the $\sigma$ resonance 
generates an unphysical pole in the physical sheet below threshold corresponding to a bound 
state at around 0.2 GeV. 
iii) The non-resonant contribution from the terms proportional to $\gamma_1$ and 
$\gamma_2$ in Eq.(\ref{3pitaylor}) are as big as the pole contribution and we see that 
the final shape of the absolute value of the amplitude is completely distorted as compared with 
the pure pole contribution. Indeed, the full results from ref.\cite{npa} show a zero at 
around 98 MeV. This is the Adler zero due to chiral
 symmetry, so that in the chiral limit the pseudoscalar interactions vanish at $s=0$. 
 At leading order in
CHPT \cite{gl}, this Adler zero sits as well at 98 MeV, very close to the position of the dip in the 
dotted and dashed-dotted lines and in good agreement with the zero in the total amplitude. 
In fact, because of the presence of this Adler zero, one 
 can understand why the background turns out to be so big. If there is a pole that affects 
 pretty much the low energy region, as in our case with the $\sigma$ resonance, 
 it is then necessary a large background to cancel the pole contribution so that 
 an Adler zero can happen.

\begin{table}
\begin{center}
\begin{tabular}{|lrrr|}
\hline
Resonance & $a_n$ & $\delta_n$ & Fraction \\
          &       & (radians) &         \\
	  \hline
NR & 0.57 & 0.30 & 12$\%$ \\
$\sigma\pi^+$ & 0.10 & 3.40 &79$\%$\\
$\rho^0(770)\pi^+$ & 1 (fixed) & 0 (fixed) & 35$\%$\\
$f_0(980)\pi^+$ & 0.47 & 2.90 & 8$\%$\\
$f_0(1370)\pi^+$ & 0.24 & 2.09 & 2$\%$\\
$f_2(1270)\pi^+$ & 0.79 & 1.05 & 22$\%$\\
$\rho^0(1450)\pi^+$ & 0.20 & 5.30 & 1$\%$\\
\hline
$\chi^2/\nu$ & 3/152& & \\
\hline
\end{tabular}
\caption{\small Results of the reproduction of the parameterization of ref.\cite{prld3pi} 
summarized in Eq.(\ref{expprld3pi}), removing the BW of the $\sigma$ for its 
 pole contribution, Eq.(\ref{purespole}). For every resonance we list the resulting 
 magnitude $a_n$ in the second column, the relative phase $\delta_n$ in radians 
 in the third column and the fraction of this decay mode in the fourth one.
\label{tab:spole}}
\end{center}
\end{table}

\begin{figure}[ht]
\psfrag{degrees}{\small Events/0.04 (GeV$^2$)}
\psfrag{MeV}{\small GeV$^2$}
\centerline{\epsfig{file=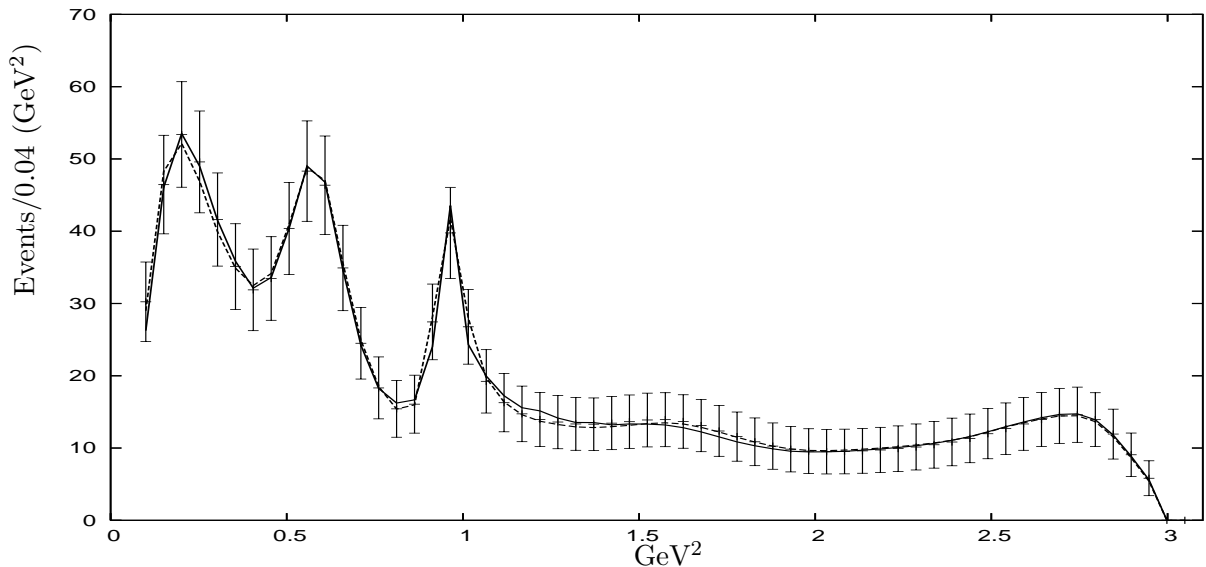,height=6.5in,width=3.0in,angle=-90}}
\vspace{0.2cm}
\caption[pilf]{\protect \small $m^2(\pi\pi)$ projections with "data" points from 
Eq.(\ref{expprld3pi}) of ref.\cite{prld3pi}. Our results are given by the dashed line with the
 pole contribution, Eq.(\ref{expprld3pi}), instead of the BW for the $\sigma$ of ref.\cite{prld3pi}.  
 The solid line corresponds to treat the S-wave FSI following Eq.(\ref{unisum3}). 
\label{fig:3pi_proy}}
\end{figure}

 Before making use of the full results of ref.\cite{npa} let us  
 substitute in Eq.(\ref{expprld3pi}) the BW contribution given 
in Eq.(\ref{bwpropa}), as employed in ref.\cite{prld3pi}, by the pure 
 $\sigma$ pole contribution,
 \be
 \frac{a_1 e^{i\delta_1}}{s-s_\sigma}~,
 \label{purespole}
 \ee
 located in the position given in Eq.(\ref{residua_s}). 
  This pole contribution, as discussed above and shown in Fig.\ref{fig:3pitaylor}a,  has 
a phase motion in agreement with that from the S-wave $I=0$ phase shifts. We keep the 
rest of terms in Eq.(\ref{expprld3pi}) and fit the $a_i$ and $\delta_i$ so as to reproduce 
the results from the parameterization employed in E791 with the values for the parameters 
given by their fit with the $\sigma$. As in ref.\cite{prld3pi}  the magnitude and 
phase of the $\rho(770)$ vector resonance parameters, $a_n$ and $\delta_n$, respectively, are fixed. 
 In that fit apart from the $a_i$ and $\delta_i$, the 
authors also leave as free parameters the mass and width of the $\sigma$. Notice that we do 
not include any Blatt-Weisskopf factors in Eq.(\ref{purespole}). In order to reproduce the results 
from ref.\cite{prld3pi} we fit a Dalitz plot with $20\times 20$ bins (a figure very standard  
in this E791 analysis) normalized to the total number of events once the background is 
subtracted, namely 1124 events. This Dalitz plot  
is generated from the parameterization employed in ref.\cite{prld3pi} for the signal only,
corresponding to Eq.(\ref{expprld3pi}). The resulting fit is very 
good with a low $\chi^2/\nu=3/152$ and the obtained magnitudes and phases are given in 
Table \ref{tab:spole}. On the other hand, we show in Fig.\ref{fig:3pi_proy} the $s_{12}$ 
projection by the dashed line, while the results from the parameterization of ref.\cite{prld3pi}
 correspond to the points.  We then conclude that  we are able to reproduce the 
results of the E791 while keeping the constraint that the phase of the $\sigma$ contribution 
follows the $I=0$ S-wave $\pi\pi$ phase shifts in a rather straightforward manner. 
An important difference of our fit in Table \ref{tab:spole} with respect to the second 
fit of \cite{prld3pi} is the clear dominance of the
 $\sigma$ pole in our case, 89$\%$, as compared with the previous reference where its fraction, 
 though also dominant, is lower, around 1/2.

Now, let us consider the full results of ref.\cite{npa}. In this reference the
partial waves are written as a product of two matrices (for the general case of 
coupled channels) as 
\be
T= \left[I+N\cdot g\right]^{-1}\cdot N~,
\label{npatmatrix}
\ee
 where $\pi\pi$ is channel 1 
and $K\bar{K}$ is channel 2, in such a way that e.g.
 $T_{11}$ is the $I=0$ elastic $\pi\pi$ S-wave amplitude and so on. The diagonal matrix $g(s)$  
corresponds to the unitarity bubbles of every channel and the matrix $N$ is determined by
 matching the previous expression with the chiral series at leading order, although the structure 
 of Eq.(\ref{npatmatrix}) is valid to any order. At lowest order 
 $N$ is the matrix of leading order CHPT amplitudes \cite{npa}. Now the important point for us is that given 
a general vertex $N_{ij}$, projected in a certain partial wave and connecting 
channels $i$ and $j$, the summation of the unitarity bubbles just implies to
multiply the matrix $N$ by the inverse of the matrix $\left[I+N\cdot g\right]$, as follows 
from Eq.(\ref{npatmatrix}). In such a way that 
if the final channel $\ell$ is produced from the initial  channel $k$ we then have:
\be
\label{unisum}
T_{\ell k}=\sum_{j}\left[I+N\cdot g\right]^{-1}_{\ell j} N_{jk} ~,
\ee
so that $j$ runs over all the possible intermediate states, that are also 
produced from channel $k$ with the original amplitude $N_{jk}$.  
 Now we want to study a completely analogous situation where channel $\ell$ is 
produced from a given production process starting with the $D^+$ meson. Then, 
if call by $\xi_j$ the original amplitude for producing channel $j$ the 
resummation of the unitarity bubbles must be given in exactly the same way 
as in Eq.(\ref{unisum}),\footnote{The analytical properties of $N_{ij}$ and $\xi_j$, in general 
terms, are the same, that is, analytical functions except for the presence of cuts and poles.} so that:
\be
\label{unisum2}
\xi_\ell \rightarrow \sum_j \left[I+N\cdot g\right]^{-1}_{\ell j} \xi_{j} ~.
\ee
In ref.\cite{npa} the $\pi\pi$ and $K\bar{K}$ channels were considered 
for the study of the meson-meson $I=0,~1$ S-waves. The $\sigma$ and 
$f_0(980)$ appear as poles in the second Riemann sheet of the resulting 
partial waves. Thus, instead of considering Eq.(\ref{expprld3pi}), that includes 
explicitly the BW's associated with these resonances \cite{prld3pi}, we will 
instead apply Eq.(\ref{unisum2}) to correct by final state interactions. In 
this way if we denote by $D=\left[I+N\cdot g\right]$, we will have:
\be
\label{unisum3}
\left[D^{-1}_{11}(s_{12})+D^{-1}_{11}(s_{13})\right] a_{\pi\pi} e^{i\delta_{\pi\pi}}+
\left[D^{-1}_{12}(s_{12})+D^{-1}_{12}(s_{13})\right]a_{K\bar{K}}e^{i\delta_{K\bar{K}}}~,
\ee
in return of the contribution for the $\sigma$ and $f_0(980)$. Notice that we do not include
 in (\ref{unisum3}) any form factor through 
Blatt-Weisskopf terms $F^{(0)}_n$ and $F_D^{(0)}$ for the two scalar resonances. 
 The rest of contributions,  NR plus the ones from the $f_0(1370)$ and vector and tensor resonances 
 in Table \ref{tab:spole} are the same as in ref.\cite{prld3pi}, although their associated
 parameters $a_n$ and $\delta_n$ are  fitted again. As in the case before, we fit a 
 Dalitz plot with $20\times 20$ bins generated from ref.\cite{prld3pi} and normalized to 
 the same number of total events. The resulting $\chi^2/\nu=2/152$ is of the same good quality 
 as before, indicating an accurate reproduction of the amplitudes of ref.\cite{prld3pi}. In Table
\ref{tab:sdmatrix} we show the values of the parameters that we have fitted 
and in Fig.\ref{fig:3pi_proy} we give by the solid line the energy projections of any neutral 
$\pi\pi$ subsystem.

\begin{table}[ht]
\begin{center}
\begin{tabular}{|lrrr|}
\hline
Resonance & $a_n$ & $\delta_n$ & Fraction \\
          &       & (radians) &         \\
	  \hline
NR & 0.70 & $-0.43$ & 17$\%$ \\
$(\pi\pi)\pi^+$ & 0.31 & 1.27 & 102$\%$\\
$(K\bar{K})\pi^+$ & 0.11 & $-0.39$ & 6$\%$\\
$\rho^+(770)\pi^+$ & 1 (fixed) & 0 (fixed) & 36$\%$\\
$f_0(1370)\pi^+$ & 0.31 & 1.99 & 3$\%$\\
$f_2(1270)\pi^+$ & 0.77 & 1.01 & 21$\%$\\
$\rho^0(1450)\pi^+$ & 0.20 & 5.48 & 1$\%$\\
\hline
$\chi^2/\nu$ & 2/152& & \\
\hline
\end{tabular}
\caption{\small Results of the reproduction of the parameterization of ref.\cite{prld3pi} 
summarized in Eq.(\ref{expprld3pi}), employing Eq.(\ref{unisum3}). For each intermediate state
 we list the resulting  magnitude $a_n$ in the second column, the relative phase $\delta_n$ in 
 radians in the third column and the fraction of this decay mode in the fourth one.
\label{tab:sdmatrix}}
\end{center}
\end{table}

As a result we see that we are able to reproduce the signal function of ref.\cite{prld3pi} rather
accurately and at the same time being able to establish that the scattering amplitudes which drive
the final state interactions corrections in $D^+\rightarrow \pi^- \pi^+\pi^+$ are the same as  
those determined from scattering data and other production processes. The same conclusion is 
obtained in ref.\cite{focus}. This reference performs a K-matrix fit to data from $D^+$ and $D_s^+$ decays to $3\pi$, 
although the resulting fits have significative lower confidence levels than those of the E791 
Collaboration that we reproduce. Regarding the $\sigma$ resonance we see that it exhibits the same
 behaviour in the D-decays as in the rest of known processes, althogh in ref.\cite{focus} is excluded. 
It is worth mentioning that the K-matrix  employed in ref.\cite{focus} does not meet the chiral requirement
 of a soft expansion for low energies  and, in particular, it does not fulfill the chiral constraints 
imposed by the chiral power counting, which requires a $\pi\pi$ scattering amplitude to start at order
 $p^2$, instead of order $p^0$ as in \cite{focus}. This rises serious doubts about the applicatibility
  of the results from the K-matrix employed in ref.\cite{focus} regarding the 
existence of the $\sigma$ resonance since meaningful
  stuctures are required in order to extrapolate the T-matrix in the energy complex plane away 
 from the physical real  axis, where it has been tested.  

Before ending this section let us discuss why the $\sigma$ resonance is more visible in 
the $D$ decays than in $\pi\pi$ scattering. One important point is the presence of the huge 
background (those terms proportional to $\gamma_1$, $\gamma_2$, etc... in Eq. (\ref{expprld3pi})) 
 in the $I=0$ S-wave $\pi\pi$ partial wave as discussed at the beginning of this 
section. There we found out that this background is required in order to preserve the Adler zero 
 in the presence of a light $\sigma$ resonance, so that a cancellation can occur close to 
threshold. In this way, the standard pole like structure of a resonance, e.g. that for the case  
the $\rho$ resonance, is completely destroyed. The question now is why this background does not happen in $D$ decays (or 
as well in $B$ decays \cite{bdecays}). The main point was 
already discussed in the second entry of ref.\cite{bdecays}, and arguments in this direction were also put 
forward in ref.\cite{bugg}. The $D$ or $B$ mesons can be identified
with pseudoscalar sources coupled directly to a pion while the other two pions can be thought to
couple just to a scalar source. As a direct application of CHPT power counting one realizes that 
this is not suppressed by any power of momentum or quark mass and then there is a priori no
 reason why the $\sigma$ meson should be screened by such large backgrounds as occur in the 
 scattering case. Indeed, this is a result from Eq.(\ref{unisum3}), and can be seen 
 by just making a Laurent expansion of the $D^{-1}$ matrix around the $\sigma$ pole and 
 then we check the absence of a significant background, contrarily to the scattering 
 case. Certainly the scalar form factor of two pseudoscalars is not renormalization
  group invariant but this just amounts to a global quark mass multiplying factor and does not
   induce any energy dependence that could distort the pole structure of the resonance, 
 contrarily to scattering where the Adler zero happens for a specific energy.

\section{FSI in the $D^+_s\rightarrow \pi^- \pi^+\pi^+$ Decay}
\label{sec:ds3pi}
\begin{figure}[ht]
\psfrag{degrees}{\small Events/0.07 (GeV$^2$)}
\psfrag{MeV}{\small GeV$^2$}
\centerline{\epsfig{file=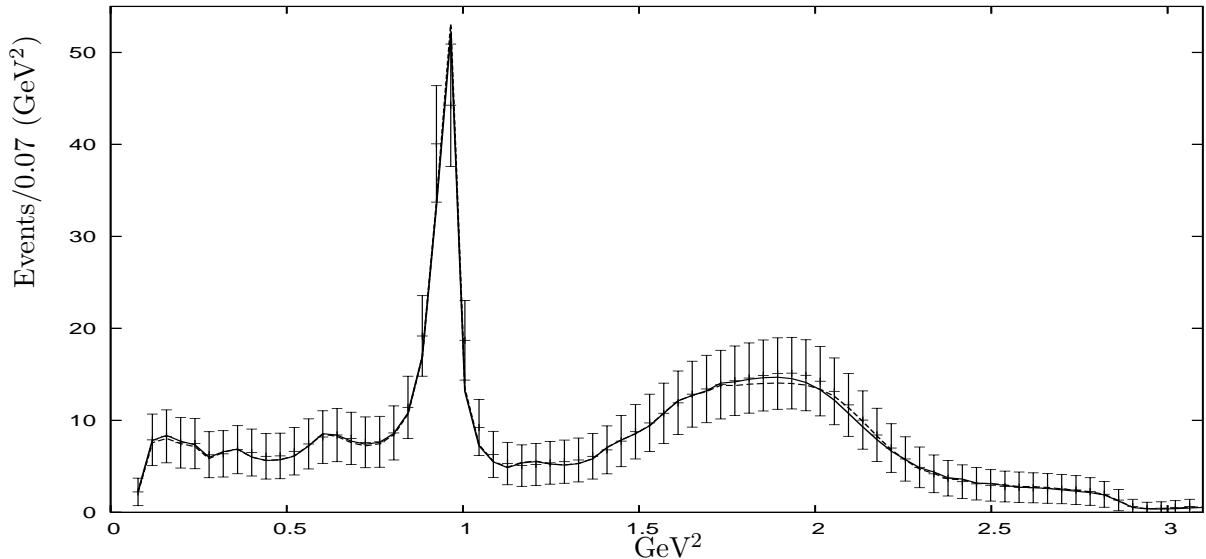,height=6.5in,width=3.0in,angle=-90}}
\vspace{0.2cm}
\caption[pilf]{\protect \small $m^2(\pi\pi)$ projections for "data"
\cite{prlds3pi}  and our results, solid and dashed lines, with the FSI 
given by Eq.(\ref{unisum2}). In the dashed line the mass and width of the $f_0(1370)$ are 
allowed to float in the fit, second and third columns of Table \ref{tab:ds3pi}. 
In the solid line the mass and width of the $f_0(1370)$ are fixed to 
the values of ref.\cite{prlds3pi}, fourth and fifth columns of Table \ref{tab:ds3pi}.
\label{fig:ds_proy}}
\end{figure}

\def\theequation{\arabic{section}.\arabic{equation}}
\setcounter{equation}{0}
In this section we consider the decay $D^+_s\rightarrow \pi^- \pi^+\pi^+$ where the 
$f_0(980)$ resonance plays a central role as clearly seen in Fig.\ref{fig:ds_proy}. 
The "data" points  correspond to the signal function of the E791 Collaboration given by 
 the analogous formula to Eq.(\ref{expprld3pi}) but applied to this other decay. The set of
 intermediate resonances considered in ref.\cite{prlds3pi} contains the $f_0(980)$, $\rho^0(770)$, $f_2(1270)$, 
 $f_0(1370)$ and the $\rho^0(1450)$. In the fit performed in ref.\cite{prlds3pi} the mass and
 couplings of the $f_0(980)$ were allowed to float with an energy dependent width given by:
 \be
 \Gamma_{f_0}(s)=g_\pi \sqrt{s/4-m_\pi^2}+g_k
 \frac{1}{2}\left(\sqrt{s/4-m_{K^+}^2}+\sqrt{s/4-m_{K^0}^2}\right)~,
 \ee
to be substituted in the expression for $BW_n(s_{12})$ in Eq.(\ref{bwpropa}). 
The striking fact is  the small value for the coupling 
$g_K$ in the fit of ref.\cite{prlds3pi}, compatible with zero,  whereas the 
coupling of $g_\pi$ is much larger. This result is very puzzled because is well known that the 
$f_0(980)$ has great affinity to couple with strangeness sources, as seen e.g. in $\phi$ decays 
\cite{fi} or in the $J\Psi\rightarrow \phi \pi\pi$ decay \cite{jpsi}.  We show 
that we can reproduce the amplitude for the signal of ref.\cite{prlds3pi} making use of the amplitudes of
ref.\cite{npa}, where the $f_0(980)$ has more standard properties regarding its couplings. 
We employ Eq.(\ref{expprld3pi}) of ref.\cite{prlds3pi} but, as in the previous section, we remove 
the $f_0(980)$ BW contribution of \cite{prlds3pi} and take into account the FSI by employing
Eq.(\ref{unisum3}), in terms of the $I=0$ S-wave $\pi\pi$ and $K\bar{K}$ states. 
In order to reproduce the amplitudes of ref.\cite{prlds3pi} we proceed as in section 
 \ref{sec:3pi}. 
For that we fit a Dalitz plot  with $20\times 20$ bins, normalized to 625 events. The Dalitz
 plot is generated from Eq.(\ref{expprld3pi}) with the parameters given in
 ref.\cite{prlds3pi}. We have performed two types of fits, given in Table \ref{tab:ds3pi}. 
 One in which the mass and width of the $f_0(1370)$ is fixed to the values of ref.\cite{prlds3pi} and 
 another in which, as in the previous reference, we let to float their values. The former 
corresponds to the second and third columns of Table \ref{tab:ds3pi} and the latter to the last 
two columns. Furthermore, in Fig.\ref{fig:ds_proy} we show the energy projection of  any 
neutral $\pi\pi$ subsystem, the solid line refers to the former fit and the dashed line 
to the latter. 
 
We see that the reproduction of the signal function of ref.\cite{prlds3pi} is very fair in both
cases. In connection with the fore mentioned unexpected values of the couplings of the $f_0(980)$ as
determined in ref.\cite{prlds3pi}, we refer the interested reader to ref.\cite{mixing} for a
detailed study of the masses and couplings of the lightest scalar resonances ($\sigma$, 
$\kappa$,  $f_0(980)$ and $a_0(980)$), showing
that they obey rather accurately a standard SU(3) analysis and constitute the lightest scalar
 nonet. Other interesting studies on the $f_0(980)$ in $D_s$ and $D$ decays are
 refs.\cite{navarra,nielsen}.

\begin{table}[H]
\begin{center}
\begin{tabular}{|l|rr||rr|}
\hline
Resonance & $a_n$ & $\delta_n$ & $a_n$ & $\delta_n$  \\
          & Fraction & (radians) & Fraction($\%$) & (radians) \\
	  \hline
NR & 0.40 & 0.16 & 0.40 & $-0.24$  \\
   &  13$\%$    &      &  14$\%$    & \\
$(\pi\pi)\pi^+$ & 0.28 & 2.36 & 0.25 & 2.23\\
&  6$\%$  & & 5$\%$& \\
$(K\bar{K})\pi^+$ & 1(fixed) & 0(fixed)& 1(fixed) & 0(fixed)\\
&  78$\%$  & & 84$\%$& \\
$\rho^+(770)\pi^+$ & 0.24 & 0.14 &  0.22 & 0.19  \\
&   4$\%$  & & 4$\%$& \\
$f_0(1370)\pi^+$ & 0.60 & 1.68 & 0.57& 2.01\\
&   28$\%$ & &28$\%$ & \\
$f_2(1270)\pi^+$ & 0.50 & 0.39 &0.48  &0.41 \\
&  20$\%$  & &19$\%$ & \\
$\rho^0(1450)\pi^+$ & 0.25 & 0.67 &0.24  &  0.87\\
&   5$\%$ & & 5$\%$& \\
\hline
$\chi^2/\nu$ & 11/142 & & 8.5/140& \\
\hline
\end{tabular}
\caption{\small Results of the reproduction of the parameterization of 
ref.\cite{prlds3pi} 
summarized in Eq.(\ref{expprld3pi}) for the $D^+_s\rightarrow \pi^-\pi^+\pi^+$. 
For each intermediate state we list the resulting 
 magnitude $a_n$ in the second column, the relative phase $\delta_n$ in radians 
 in the third column and the fraction of this decay mode just below the value of 
 every $a_n$, analogously for the fourth and fifth columns. The second and third column correspond to the fit with the mass and 
 width of the $f_0(1370)$ fixed to the values of ref.\cite{prlds3pi}, while in the fourth 
 and fifth columns these values are allowed to float in the fit. In the latter case, 
 $m_{f_0(1370)}=1.46$ and $\Gamma_{f_0(1370)}=0.16$ GeV.
\label{tab:ds3pi}}
\end{center}
\end{table}

\section{FSI in the $D^+\rightarrow K^- \pi^+\pi^+$ Decay}
\label{sec:dk2pi}
\begin{figure}[ht]
\psfrag{degrees}{\small Phase shifts (degrees)}
\psfrag{gev}{\small GeV}
\psfrag{abs}{\small (Absolute value)$^2$}
\psfrag{Kpi}{\small $K\pi$ $I=1/2$ S-wave}
\centerline{\epsfig{file=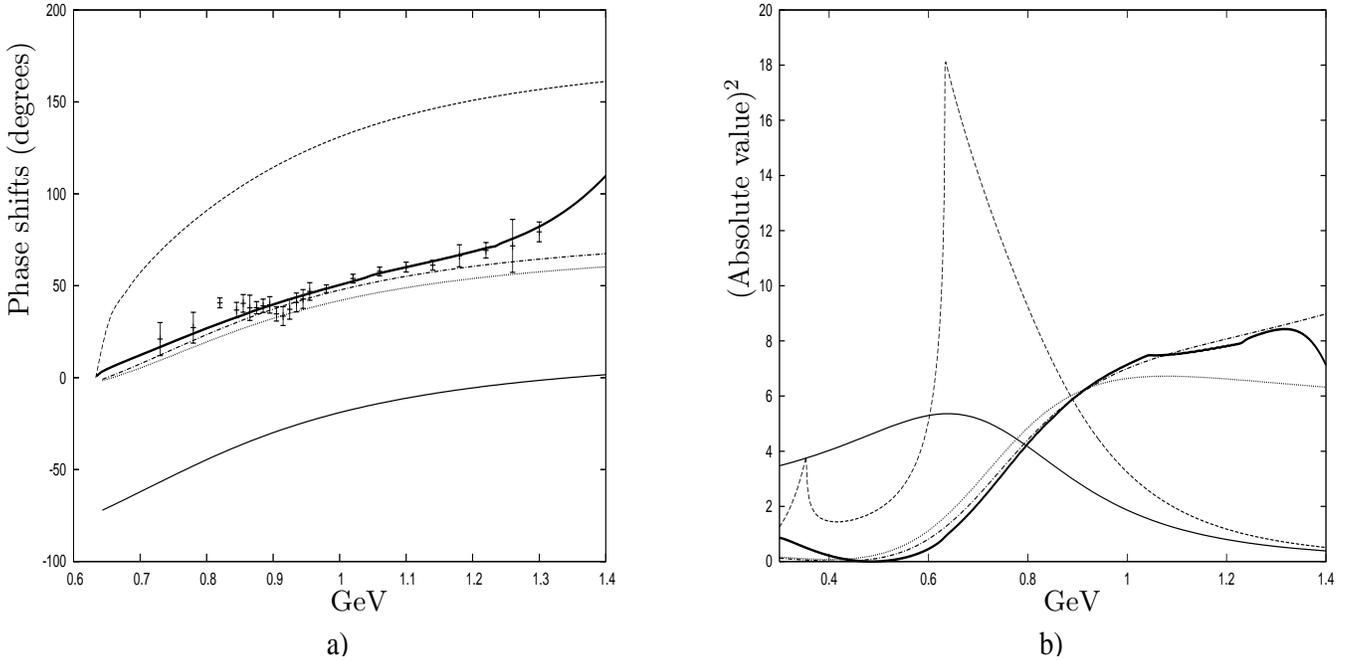,height=3.5in,width=7.0in,angle=0}}
\vspace{0.2cm}
\caption[pilf]{\protect \small S-wave $I=1/2$ $K\pi$ phase shifts (left panel) 
and modulus square of this partial wave, normalized such that the residue at 
the $\kappa$ pole is one (right panel).
The data points are from  ref.\cite{estabrooks}. 
The dashed lines correspond to the  BW for the $\kappa$ 
resonance employed in ref.\cite{prldk2pi}. 
The solid lines correspond to the pure $\kappa$ pole contribution from Eq.(\ref{kpitaylor}). 
The results from the contribution 
of the $\kappa$ pole plus the first non-resonant term in 
Eq.(\ref{kpitaylor}) are shown by the dotted lines. Finally, the 
dashed-dotted lines are the results from the Laurent expansion of 
Eq.(\ref{kpitaylor}) keeping all the terms shown.
\label{fig:kpitaylor}}
\end{figure}

\def\theequation{\arabic{section}.\arabic{equation}}
\setcounter{equation}{0}
The study of the high statistics Dalitz plot of the decay 
$D^+\rightarrow K^- \pi^+\pi^+$, with a sample of 15090 and an estimated 
 background of a 6$\%$, was performed in ref.\cite{prldk2pi} within the 
E791 Collaboration in Fermilab. 
 In order to introduce the controversial situation regarding the amplitude employed to describe data in ref.\cite{prldk2pi}, let us
briefly discussed the fitting process followed in this reference. When in this reference the amplitude, 
 following  the isobar model, is written as the coherent sum of well established resonances, as so
 quoted in the Particle Data Group ref.\cite{pdg04}, the  description of the fit is poor.
  Indeed the $\chi^2$ per degree of freedom found in ref.\cite{prldk2pi} is $\chi^2/\nu=167/63$. The main 
 discrepancies between data and the parametrized amplitude come, particularly, from the 
 low energy region, 
below 0.6 GeV$^2$, and at higher energies at around 2.5 GeV$^2$. 
In order to improve the quality of the
fit of the Dalitz plot the authors of ref.\cite{prldk2pi} allowed 
the mass and width of the scalar resonance $K^*_0(1430)$ to float in the fit and also included 
Gaussian scalar form factors a la Tornqvist, with the meson radii as additional free parameters.
 Despite that the $\chi^2/\nu$ is also around 2. Finally, the authors of ref.\cite{prldk2pi}, together with the previous additions, also 
included the $\kappa$ resonance, referred as $K^*_0(800)$ in the last edition of the 
PDG \cite{pdg04} but qualified as controversial. The quality of the fit substantially 
improves and the $\chi^2/\nu=46/63$. However, the situation is far from being resolved 
since: i) the width of the well known $K^*_0(1430)$,
 allowed to float as discussed above, turned out to be a factor of two lower than its 
clearly determined value in scattering experiments and other production processes. ii) 
The Breit-Wigner employed to describe the $\kappa$ resonance does not
 follow the elastic S-wave $I=1/2$ $K\pi$ phase shifts. This is shown in 
 Fig.\ref{fig:kpitaylor} where the points are the elastic S-wave $I=1/2$ $K\pi$ phase shifts 
from \cite{estabrooks} and the dashed line corresponds to the phase of the relativistic BW 
of the $\kappa$. The discrepancy 
 is manifest, specially close to threshold where the variation for the phase of the BW is much 
 faster than that corresponding to data. Let us remember that the S-wave $I=1/2$ $K\pi$ scattering 
 is elastic below around 1.3 GeV, since both 
the $K\eta$ and $K\pi\pi$ channels are negligible below that energy, as clearly shown 
both experimentally \cite{estabrooks,aston} and theoretically \cite{jamin}, and then only the 
$K\pi$ elastic channel matters in this energy region and the discrepancy cannot come 
due to inelastic effects. 

Denoting by 1 the $K^-$ and by 2 and 3 the equal pions, the amplitudes {\it employed} 
in ref.\cite{prldk2pi} can be written as:
\be
{\cal A}=a_0 e^{i\delta_0} {\cal N}_0+\sum_{n=1}^N a_n e^{i\delta_n} {\cal
A}_n(s_{12},s_{13}) {\cal N}_n~.
\label{expprldk2pi}
\ee
The different ingredients contained in the previous equation were discussed in detail 
in section \ref{sec:3pi}. Here we only mention that in ref.\cite{prldk2pi} the BW expression 
(\ref{bwpropa}) has opposite sign.

The set of resonances that exchange in Eq.(\ref{expprldk2pi}) contains the
$K^*_0(892)$, $K^*_0(1430)$, $K_2^*(1430)$, $K^*(1680)$ and, in the 
parameterization that reproduces faithfully the experimental data, the $\kappa$
 or $K^*_0(800)$. As discussed above the most controversial aspects present in 
 Eq.(\ref{expprldk2pi}) involves the $I=1/2$ S-wave $K\pi$ partial wave.

Now, we want to show that we are able to reproduce the amplitude 
given in Eq.(\ref{expprldk2pi}) as employed in ref.\cite{prldk2pi} but 
using the S-wave $I=1/2$ $K\pi$, $K\eta$ and $K\eta'$ coupled channel partial 
waves derived in ref.\cite{jamin}. These T-matrices are obtained from 
Chiral Perturbation Theory (CHPT) \cite{wein,gl} at next-to-leading order, supplemented with 
the exchange of explicit resonance fields in a chiral invariant manner 
\cite{rafael}, plus a unitarization scheme compatible with the chiral 
expansion (chiral unitariy approach). Furthermore, additional 
constrains from large $N_c$ QCD are considered as well in order to restrict the 
number of free parameters, for further details we refer to
refs.\cite{jamin,jaminff}. These T-matrices are able to provide an accurate 
reproduction of the $K\pi$ S-wave amplitude and of the $I=1/2$ and $I=3/2$ S-wave phase 
shifts \cite{estabrooks,aston} up to around 2 GeV. Later on, they were
employed for calculating through dispersion relations the 
strangeness changing scalar form factors \cite{jaminff}, the 
light quark masses within QCD scalar sum rules  \cite{jaminms} and a crucial 
counterterm needed in the present and precise studies of $K_{\ell 3}$ decays 
\cite{jaminckm}.

The S-wave $I=1/2$ T-matrices from ref.\cite{jamin} contains three poles in the
appropriate Riemann sheets corresponding to the $\kappa$, $K^*_0(1430)$ and 
 the $K^*_0(1950)$. Their pole positions  are 
around $708-i\, 305$, $1450-i\,142$ and $1910-i\,27$ MeV, 
respectively.\footnote{These pole positions
vary slightly depending of the fit taken from ref.\cite{jamin}, although we have
presented the values from the so called preferred fit, which behaves in
the most appropriate way when considering scalar form factors and QCD sum rules.}  Let us
 concentrate on the first two resonances
 which are those that are relevant for the $D^+$ decay into $K^-\pi^+\pi^+$, since
the maximum value of the total center of mass energy of the subsystems (12) 
or (13) is $m_{D^+}-m_{\pi^+}= 1.73$ GeV, clearly below the influence of 
the $K^*_0(1950)$ pole. Let us remark that the given pole positions are obtained as a result
 of the full T-matrices derived in ref.\cite{prldk2pi}, which are not just any sum of  pole contributions,
although these can dominate in some energy regions. 
The value for the width of the well established and clearly seen $K^*_0(1430)$ resonance 
is perfectly compatible with that quoted by the PDG \cite{pdg04}, 
$294\pm 23$ MeV,\footnote{Remember that the $double$ of minus the
imaginary part of the pole position corresponds to the width of a resonance.
This is indeed a possible and unambiguous definition of the width of a 
resonance.} while that from the fit of the E791 Collaboration \cite{prldk2pi}, 
$175\pm 12 \pm 12$ MeV, is almost a factor of two lower. In addition, the mass obtained 
in ref.\cite{prldk2pi} for the $K^*_0(1430)$ is out of the range given in 
the PDG as well. We also mention that the width of the $\kappa$ given by the 
E791 Collaboration, around 400 MeV, is substantially lower than the value from the pole
 position from ref.\cite{jamin}, around 600 MeV.

Due to the already mentioned discrepancy between the phase motion in energy of the 
BW for the $\kappa$ from the E791 Collaboration, let us perform a Laurent series around 
the $\kappa$ pole position of the $I=1/2$ $K\pi$ S-wave amplitude,
 \be
 t_{11}=\frac{\gamma_0^2}{s-s_\kappa}+\gamma_1+\gamma_2(s-s_\kappa)+\ldots~.
 \label{kpitaylor} 
 \ee
 with the pole position and residua obtained from ref.\cite{jamin}, with the 
 values:
 \ba
 s_\kappa&=&(0.71-i 0.31)^2 \hbox{~GeV}^2~,~\gamma_0^2=19.0+i 5.9 \hbox{~GeV}^2~,\nn\\
 \gamma_1&=&12.5+i 43.1 ~,~\gamma_2=0.8+i7.3 \hbox{~GeV}^{-2}~.
 \label{taylork}
 \ea
 In  Fig.\ref{fig:kpitaylor} we show in the left and right panels the phase and 
normalized absolute value of this partial wave, respectively. 
The thinner solid lines correspond to the pure pole contribution in Eq.(\ref{kpitaylor}).  
We see in Fig.\ref{fig:kpitaylor}a that the phase of this pole contribution does not vanish
 at threshold, but  
runs parallel to the experimental phase shifts, so that the difference with respect to them 
 keeps constant along energy up to the increase in the last points due to the closeness 
 of the $K^*_0(1430)$ and the opening of the $K\eta'$ channel. Thus 
the phase of the pure pole contribution from Eq.(\ref{kpitaylor}) does follow the motion of
 the experimental S-wave $I=1/2$ $K\pi$ phase shifts, in contrast with the BW phase indicated 
 by the dashed line. It is also 
 worth realizing that the phase of the pure $\kappa$ pole contribution starts at $-90$ 
 degrees at threshold and this is the reason why its value is not $+90$ degrees at the mass 
 of the $\kappa$ resonance, $708$ MeV, but happens much later. By the same reason, 
 for $s\rightarrow \infty$ one gets only $+90$ degrees from this pole contribution. 
 The agreement between the experimental 
phase shifts and the expansion (\ref{kpitaylor}) is reached rather fast. The dotted lines correspond 
to keep as well the $\gamma_1$ term while the dashed-dotted ones correspond to keep all the terms
shown in this equation. The thick solid lines are the full results from ref.\cite{jamin}. In 
Fig.\ref{fig:kpitaylor}b we consider the absolute value of the partial wave under consideration, 
but normalized such that the residue at the pole position is one, so that we divide 
by $\gamma_0^2$ the series in Eq.(\ref{kpitaylor}). The source of each line is the same 
as already explained and let us notice that this figure starts at around 0.3 GeV, below 
threshold.  We want to stress three important facts: i) 
The very different behaviour of the $\kappa$ BW employed by the E791 Collaboration and 
the absolute value from the partial wave of ref.\cite{jamin}. ii) The first peak in the BW. This 
peaks results because the BW formula (\ref{bwpropa}) for the $\kappa$ resonance 
generates an unphysical pole in the physical sheet below threshold at 
$0.46 \pm i 0.11$ GeV, although it has a negligible impact above threshold. 
iii) The non-resonant contributions proportional to $\gamma_1$ and 
$\gamma_2$ in Eq.(\ref{kpitaylor}) are as big as the pole contribution, and we see that 
the final shape of the absolute value of the amplitude is completely distorted as compared with 
the pure pole contribution. Indeed they almost 
cancel each other at around 0.48 GeV giving rise to a zero in the full amplitude, as clearly seen 
in the figure. This is an Adler zero  due to chiral
 symmetry, so that in the chiral limit the pseudoscalar interactions vanish at $s=0$. 
 At leading order in
CHPT \cite{gl}, this Adler zero sits at  0.48 GeV, very close to the position of the dip in the 
dotted and dashed-dotted lines and even closer to the zero in the total amplitude which also 
occurs at around 0.48 GeV.  In fact, because of the presence of this Adler zero one 
 can understand why the background turns out to be so big, in the same way as explained 
 in section \ref{sec:3pi} for the $\sigma$ case. That is, if there is a pole that largely affects 
 the low energy region, it is then necessary a large background to cancel the pole 
 contribution so that the Adler zero can occur.

\begin{figure}[ht]
\psfrag{degrees}{\small Events/0.04 (GeV$^2$)}
\psfrag{MeV}{\small GeV$^2$}
\centerline{\epsfig{file=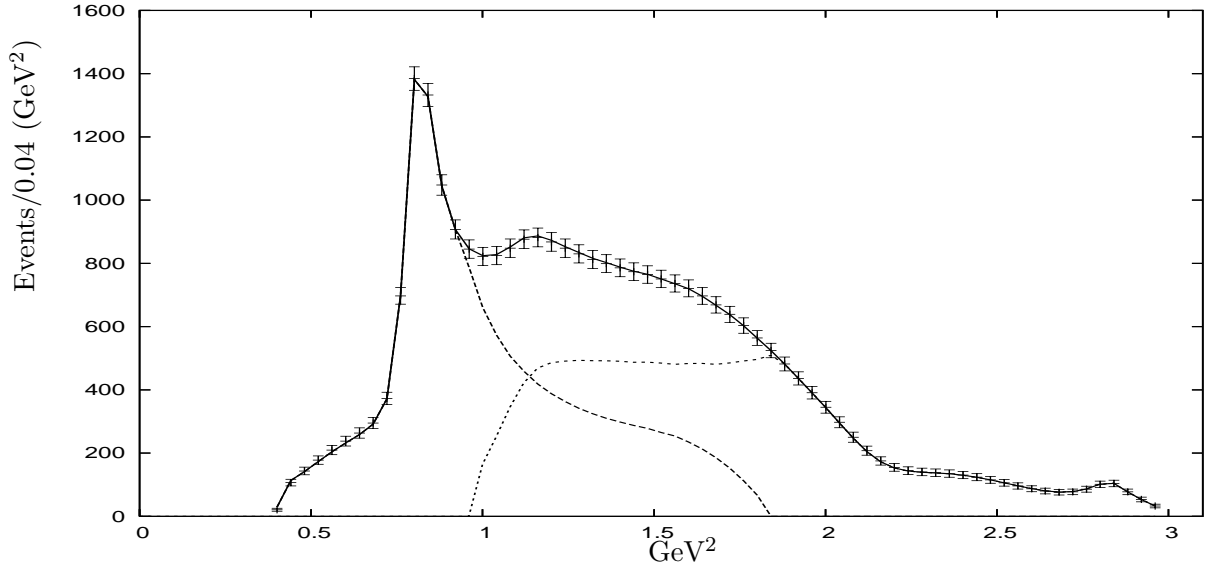,height=6.5in,width=3.0in,angle=-90}}
\vspace{0.2cm}
\caption[pilf]{\protect \small $m^2(K\pi)$ projections for "data"
\cite{prldk2pi}  and our results with the pole contribution in Eq.(\ref{expprldk2pi}) 
instead of the BW for the $\kappa$ of ref.\cite{prldk2pi}. The solid, dashed and 
dotted lines correspond 
to our results for the projections $m^2(K\pi)$, $m^2(K\pi)_{low}$ and  $m^2(K\pi)_{high}$, 
in order.
\label{fig:i12_kpole_proy}}
\end{figure}

 Before making use of the full results of ref.\cite{jamin} let us  
 substitute in Eq.(\ref{expprldk2pi}) the BW contribution given 
in Eq.(\ref{bwpropa}), as employed in ref.\cite{prldk2pi}, by the 
 $\kappa$ pole contribution,
 \be
 \frac{a_1 e^{i\delta_1}}{s-s_\kappa}~,
 \label{purekpole}
 \ee
 located in the position given in Eq.(\ref{taylork}). 
 This pole contribution, as discussed above and shown in Fig.\ref{fig:kpitaylor}a,  has 
a phase motion in agreement with that from the S-wave $I=1/2$ phase shifts. We keep the 
rest of terms in Eq.(\ref{expprldk2pi}) and fit the $a_i$ and $\delta_i$ so as to reproduce 
the results from the parameterization employed in E791, Eq.(\ref{expprldk2pi}), 
with the values for the parameters 
given by their fit C. As in ref.\cite{prldk2pi}, the magnitude and 
phase of the $K^*_0(892)$ vector resonance parameters, $a_n$ and $\delta_n$, are fixed to 
1 and 0, respectively. 
 In that fit, apart from the $a_i$ and $\delta_i$, the 
authors also leave as free parameters the masses and widths of the $\kappa$ and
 $K^*_0(1430)$, as well as the meson radii that appear in the form factors. Notice that
  we do not include either any Blatt-Weisskopf factors nor Gaussian form factors in 
  Eq.(\ref{purekpole}). In order to
   reproduce the results from ref.\cite{prldk2pi} we fit a Dalitz plot with 20$\times $20 bins
    normalized to the total number of events once the background is subtracted, 
namely 28369 events (the charge conjugate decay is also included). This Dalitz plot  
is generated from the parameterization employed in ref.\cite{prld3pi} for the signal function,
corresponding to Eq.(\ref{expprldk2pi}). The resulting fit is very 
good with a low $\chi^2/\nu=6.5/132$. The values of the resulting magnitudes and phases are given in 
Table \ref{tab:spole}.  On the other hand, we show in Fig.\ref{fig:i12_kpole_proy} the $s_{12}$ 
projection by the solid line, while the results from the parameterization of ref.\cite{prldk2pi}
 correspond to the points. The dashed line is the so called high-projection ($s_{12}>s_{13}$) and the dotted 
line corresponds to the low projection ($s_{12}<s_{13}$).  We then conclude that we are able 
to reproduce the results of the E791 at the same time that the phase of the 
$\kappa$ contribution follows the $I=1/2$ S-wave $K\pi$ phase shifts.
\begin{table}[ht]
\begin{center}
\begin{tabular}{|lrrr|}
\hline
Resonance & $a_n$ & $\delta_n$ & Fraction \\
          &       & (radians) &         \\
	  \hline
NR & 2.10 & $-5.95$ & 53.3$\%$ \\
$\kappa\pi^+$ & 0.30 & $-0.63$ & 56.7$\%$\\
$K^*_0(1430)\pi^+$ & 1.00 & 0.92 & 12.2$\%$\\
$K^*_1(892)\pi^+$ & 1 (fixed) & 0 (fixed) & 12.1$\%$\\
$K^*_2(1430)\pi^+$ & 0.17 & $-0.76$ & 0.4$\%$\\
$K^*_1(1680)\pi^+$ & 0.45 & 0.58 & 2.5$\%$\\
\hline
$\chi^2/\nu$ & 6.5/132 & & \\
\hline
\end{tabular}
\caption{\small Results of the reproduction of the parameterization of ref.\cite{prldk2pi} 
summarized in Eq.(\ref{expprldk2pi}), removing the BW of the $\kappa$ by its 
 pole contribution, Eq.(\ref{purekpole}). For each resonance we list the resulting 
 magnitude $a_n$ in the second column, the relative phase $\delta_n$ in radians 
 in the third column and the fraction of this decay mode in the fourth one.
\label{tab:kpole}}
\end{center}
\end{table}

We notice that in Table \ref{tab:kpole}  the so called NR contribution plays a  
more important role, it has a fraction of around $50\%$ while in fit C of 
the E791 Collaboration  amounts to just a  13$\%$.

\begin{figure}[ht]
\psfrag{degrees}{\small Events/0.04 (GeV$^2$)}
\psfrag{MeV}{\small GeV$^2$}
\centerline{\epsfig{file=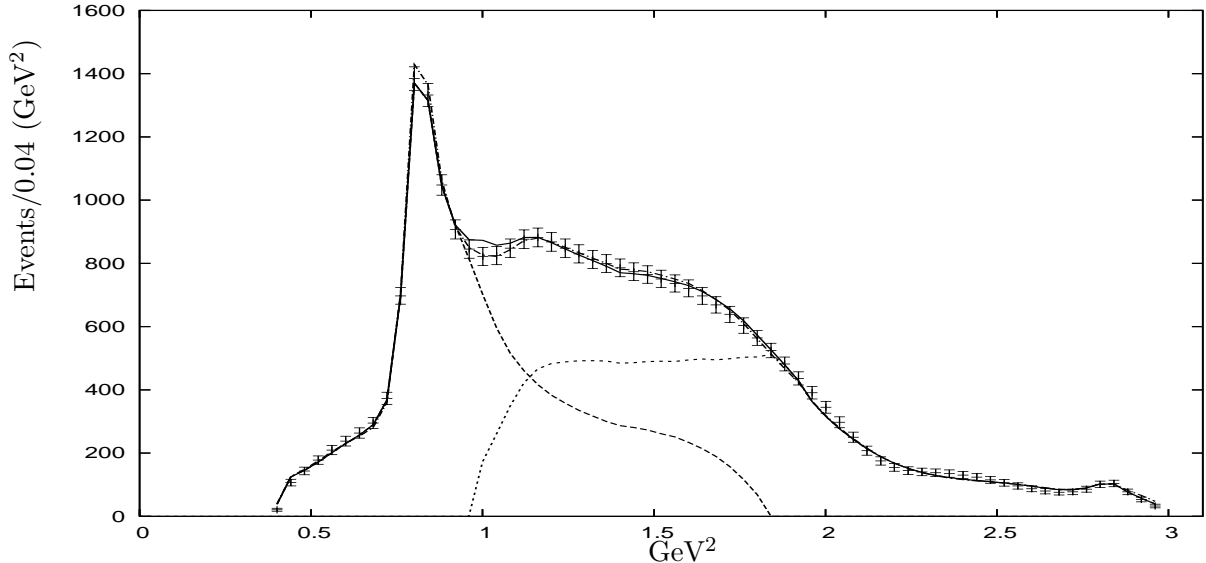,height=6.5in,width=3.0in,angle=-90}}
\vspace{0.2cm}
\caption[pilf]{\protect \small $m^2(K\pi)$ projections for "data"
\cite{prldk2pi}  and our results taking into account Eq.(\ref{unisum3b}). The solid, dashed and 
dotted lines correspond 
to our results for the projections $m^2(K\pi)$, $m^2(K\pi)_{low}$ and  $m^2(K\pi)_{high}$, 
in order. The dashed-dotted line refers to the fit when the $(K\eta)\pi^+$ channel is excluded, 
Eq.(\ref{unisum4}).
\label{fig:kpi_dmatrix_proy}}
\end{figure}

Now, let us consider the full results of ref.\cite{jamin} in order to take into account the 
FSI employing Eq.(\ref{unisum3}). In ref.\cite{jamin} the
partial waves are written as a product of two matrices (for the general case of 
coupled channel) as $T= \left[I+N\cdot g\right]^{-1}\cdot N$ such that 
e.g. $T_{11}$ is the $I=1/2$ 
elastic $K\pi$ S-wave. The diagonal $g(s)$ matrix corresponds to the unitarity bubbles 
of each channel and the matrix $N$ is determined by matching the previous
expression with the chiral series of CHPT plus resonances in the $U(3)$ case, 
for more details on this respect see ref.\cite{jamin}. In this reference the $K\pi$, 
$K\eta$ and $K\eta'$ channels are considered, hence we rewrite Eq.(\ref{unisum3}) for this 
particular case as,
\ba
\label{unisum3b}
&&\left[D^{-1}_{11}(s_{12})+D^{-1}_{11}(s_{13})\right]a_{K\pi}\, e^{i\delta_{K\pi}}
+\left[D^{-1}_{12}(s_{12})+D^{-1}_{12}(s_{13})\right]a_{K\eta}\,e^{i\delta_{K\eta}}\nn\\
&&+\left[D^{-1}_{13}(s_{12})+D^{-1}_{13}(s_{13})\right]a_{K\eta'}\,e^{i\delta_{K\eta'}}~.
\ea
Let us stress that, since the $\kappa$ and 
$K^*_0(1430)$ appear as poles in the second Riemann sheet of the partial waves of 
ref.\cite{jamin}, they are already 
taken into account  when considering Eq.(\ref{unisum3b}). Notice that we do not include 
 in Eq.(\ref{unisum3b})  for the two scalar resonances either any from factor as done in 
Eq.(\ref{expprldk2pi}) through the Blatt-Weisskopf terms $F^{(0)}_n$
 and $F_D^{(0)}$ nor Gaussian form factors. Furthermore, 
 the $K^*_0(1430)$ resonance from ref.\cite{jamin} has a mass and a width in complete 
 agreement with the values in the PDG \cite{pdg04}, which are clearly determined from the
  study of $K\pi$ scattering \cite{estabrooks,aston}. The rest of contributions present 
  in Eq.(\ref{expprldk2pi}), namely, the NR plus the ones from the vector and 
tensor resonances in Table \ref{tab:kpole}
are kept, although the parameters $a_n$ and $\delta_n$ are fit again so as  to 
reproduce the results of the signal parameterization of the E791 Collaboration.
As in the cases before, we fit a Dalitz plot with $20\times 20$ bins generated 
from ref.\cite{prldk2pi} and normalized to the total number of events. 
The resulting $\chi^2/\nu=127/128$ is acceptable, indicating a 
rather  good reproduction of the amplitudes of ref.\cite{prldk2pi}. In Table
\ref{tab:kdmatrix} we show the values of the parameters that we have fitted 
and in Fig.\ref{fig:kpi_dmatrix_proy} we show the energy projections of any neutral 
$K\pi$ subsystem.

\begin{table}[H]
\begin{center}
\begin{tabular}{|lrrr|}
\hline
Resonance & $a_n$ & $\delta_n$ & Fraction \\
          &       & (radians) &         \\
	  \hline
NR & 1.60 & 0.10 & 29.6$\%$ \\
$(K\pi)\pi^+$ & 1.66 & 4.10 & 31.8$\%$\\
$(K\eta)\pi^+$ & 0.86 & 2.63 & 2.0$\%$\\
$(K\eta')\pi^+$ & 2.33 & $-1.54$ & 9.8$\%$\\
$K^*_1(892)\pi^+$ & 1 (fixed) & 0 (fixed) & 11.6$\%$\\
$K^*_2(1430)\pi^+$ & 0.11 & $-0.62$ & 0.2$\%$\\
$K^*_1(1680)\pi^+$ & 0.72 & 0.80 & 5.9$\%$\\
\hline
 $\chi^2/\nu$ & 127/128 & & \\
\hline
\end{tabular}
\caption{\small Results of the reproduction of the parameterization of ref.\cite{prldk2pi} 
summarized in Eq.(\ref{expprldk2pi}), employing Eq.(\ref{unisum3b}). For each intermediate state we list the resulting 
 magnitude $a_n$ in the second column, the relative phase $\delta_n$ in radians 
 in the third column and the fraction of this decay mode in the fourth one. For this fit 
 we also leave as free parameters the poorly measured mass and width of the $K^*_1(1680)$, their
 values are given in the text. 
\label{tab:kdmatrix}}
\end{center}
\end{table}

As a result we see that we are able to reproduce the signal function of ref.\cite{prldk2pi} rather
accurately and, at the same time, we are able to establish that the scattering amplitudes 
driving the final state interactions corrections in $D^+\rightarrow K^- \pi^+\pi^+$ are 
compatible with those determined from scattering data \cite{estabrooks,aston}. In particular, 
the $\kappa$  resonance is present in both and the mass and 
width of the $K^*_0(1430)$ does neither differ in both cases. On
the other hand in the previous fit we have allowed to float the mass and width of the
 $K^*_1(1680)$, since they are poorly measured. In the PDG \cite{pdg04} the values reported 
 are  between 1.7 to 1.8 MeV for the mass 
and from 0.17 to 0.40 GeV for the width. The values that we obtain in the fit to the 
parameterization of ref.\cite{prldk2pi}, which employs the central values given in the 
PDG, are 
$m=1.7$ GeV and width $\Gamma=0.17$ GeV. 

Although we have already included  the $K\eta$ 
channel in the fit of Table \ref{tab:kdmatrix}, since this channel has little effect 
on the $K\pi$ scattering we then check the stability of this fit by 
removing the  $K\eta$ channel. Then, instead of Eq.(\ref{unisum3}) we now consider,
\be
\label{unisum4}
\left[D^{-1}_{11}(s_{12})+D^{-1}_{11}(s_{13})\right]a_{K\pi} e^{i\delta_{K\pi}}
+\left[D^{-1}_{13}(s_{12})+D^{-1}_{13}(s_{13})\right]a_{K\eta'}e^{i\delta_{K\eta'}}~.
\ee
The resulting $\chi^2/\nu=144/130$ is a bit larger than the previous one,  and the 
values of the parameters in the fit are rather similar so that we refrain from presenting 
them. Noneteless, we show the event projection in Fig.\ref{fig:kpi_dmatrix_proy} for this case by 
the dashed-dotted line.  On the other hand, the resulting mass and width of the $K^*(1680)$  are 1.7 GeV 
and 0.13 GeV, in order. Thus, we conclude that our results are rather stable 
under the presence or removal of the $K\eta$ channel although the resulting $\chi^2$ is somewhat 
lower when this channel is considered as well.

Finally, the same reasons advocated at the end of section \ref{sec:3pi} to explain why 
the $\sigma$ meson is clearly visible in $D^+$ decays to three pions 
can be also applied here for the $\kappa$ in the $D^+\rightarrow \pi^-\pi^+\pi^+$ decay.\footnote{As explained at the end of section \ref{sec:3pi}, but applied 
 now to our case, if we make a 
 Laurent expansion of $D^{-1}$ in Eq.(\ref{unisum3b}) around the $ \kappa$ pole one 
 checks that there is no a significant background and $D^{-1}$ is driven by the 
 $\kappa$ pole contribution around threshold up to about 1 GeV when 
  the influence of the $K\eta'$ channel and 
that of the $K^*_0(1430)$ resonance starts. Thus, the $\kappa$ resonance pole structure is not 
 distorted as happens with scattering.}

\section{Conclusions}
\label{sec:con}
\def\theequation{\arabic{section}.\arabic{equation}}
\setcounter{equation}{0}

We have considered in detail the FSI of the $D^+\rightarrow \pi^-\pi^+\pi^+$, 
$D^+_s\rightarrow \pi^-\pi^+\pi^+$ and $D^+\rightarrow K^- \pi^+\pi^+$ driven 
by the S-waves. The Dalitz plots associated with these decays were studied 
originally in refs.\cite{prld3pi,prlds3pi,prldk2pi}  showing by the first time statistically 
significant evidences for the existence of the $\sigma$ and $\kappa$ resonances. Here 
 we have payed special 
attention to those aspects still controversial after those works with the aim of improving the 
theoretical basis of the parameterizations employed in these references. In particular,
we have shown that the meson-meson S-waves with $I=0,~ 1/2$
 that drive the corresponding FSI in the previous decays  are compatible
 with those amplitudes determined from studies of scattering 
data \cite{npa,nd,jamin}, also tested in many other production 
 processes, e.g. \cite{gama,fi,jpsi,bdecays}. In particular, we have shown that
 the phase motion of the low energy FSI in the  
 $D^+\rightarrow \pi^-\pi^+\pi^+$ and $D^+\rightarrow K^- \pi^+\pi^+$ decays follows
 the elastic $I=0$ $\pi\pi$ and $I=1/2$ $K\pi$ S-waves phase shifts, in order, and how this is
compatible with the presence of the $\sigma$ and $\kappa$ resonances, respectively. We have seen that the reason for the disagreement between 
the phases associated with these resonances in the studies of
refs.\cite{prld3pi,prldk2pi} and the experimental phase shifts is the employment by
these authors of Breit-Wigner propagators. Once the BW propagator is substituted
by the pure pole contributions from Eqs.(\ref{3pitaylor}) and (\ref{kpitaylor}), for
the $\sigma$ and $\kappa$, respectively, the agreement is restored. Indeed, we have
 also shown that these pole contributions are not affected by significant
 backgrounds in the expansion of $D^{-1}$, see Eqs.(\ref{unisum3}) and
 (\ref{unisum3b}), while huge destructive backgrounds are present in the scattering.
 In this way the pole contributions are not distorted in the studied $D_s$ and $D$ decays while
 they are so in the scattering, as shown in Figs.\ref{fig:3pitaylor} and 
 \ref{fig:kpitaylor}. Furthermore, we have also considered the full results for the
S-wave FSI from refs.\cite{npa,jamin} from Eqs.(\ref{unisum3}) and
(\ref{unisum3b}). In this way we have considered simultaneously the interplay between
the $\sigma$ and the $f_0(980)$ for $I=0$ (\ref{unisum3}), and that between 
the $\kappa$ and $K^*_0(1430)$ for  $I=1/2$ (\ref{unisum3b}). This is also
important since in ref.\cite{prlds3pi} the properties of the
couplings of the $f_0(980)$ are astonishing while in ref.\cite{prldk2pi} the mass 
of the $K^*_0(1430)$ resonance is out of the range given in the PDG \cite{pdg04} and 
its width is almost a factor 1/2 smaller. Thus, our results show that one can 
also understand these FSI with "standard" $f_0(980)$ and $K^*_0(1430)$ properties.

\vspace{1cm}
\noindent {\bf Acknowledgments}

I would like to acknowledge Ignacio Bediaga for fruitful and stimulating 
 discussions, 
together with a long exchange of emails. I would also 
like to thank Isabel Guillam\'on who participated in an early stage of this research. 
I also thank Alberto C. dos Rios and Carla G\"obel for their kind disposition 
to share with me their work. Financial support from the CICYT (Spain) Grants No. FPA2002-03265 and 
FPA2004-03470 and from the EU-RTN Programme "EURIDICE", Contract No. HPRN-CT-2002-00311, is acknowledged.

\end{document}